\documentclass[showpacs,showkeys,twocolumn,prl,superscriptaddress]{revtex4-1}%
\pdfoutput=1
\usepackage{amsmath,amsfonts,amscd,revsymb,latexsym,enumerate}
\usepackage{times}
\usepackage{mathtools}
\usepackage{amssymb}

\usepackage{import}
\usepackage{xifthen}
\usepackage{pdfpages}
\usepackage{transparent}

\usepackage[caption=false,position=b]{subfig}
\usepackage[colorlinks]{hyperref}
\usepackage[capitalize]{cleveref}

\providecommand{\U}[1]{\protect\rule{.1in}{.1in}}
\providecommand{\U}[1]{\protect\rule{.1in}{.1in}}

\newcommand{\pardash}[1]{\paragraph{#1---}}

\makeatletter
\AtBeginDocument{\let\LS@rot\@undefined}
\makeatother
\begin{document}
\title{Right-sizing fluxonium against charge noise}

\author{Ari Mizel and Yariv Yanay}
\affiliation{Laboratory for Physical Sciences, College Park, Maryland 20740, USA}

\begin{abstract}

We analyze the charge-noise induced coherence time $T_2$ of the fluxonium qubit as a function of the number of array junctions in the device,  $N$. 
The pure dephasing rate decreases with $N$, but we find that the relaxation rate increases, so $T_2$ achieves an optimum as a function of $N$.  This optimum can be much smaller than the number typically chosen in experiments, yielding a route to improved fluxonium coherence and simplified device fabrication  at the same time.
\end{abstract}
\maketitle
 
\pardash{Introduction}One of the earliest superconducting qubits was the Cooper-pair box \cite{Schnirman97,Bouchiat_1998,NakamuraNat99}.  This qubit design is intuitive, with the $\left| 0 \right>$ and $\left| 1 \right>$ states corresponding physically to an excess Cooper-pair residing on or off a small superconducting island.  However, it did not take long before experiments demonstrated its acute vulnerability to ambient charge noise \cite{NakamuraPRL02}.  
One popular modification of the Cooper-pair box is the transmon qubit \cite{KochPRA2007}, which adds a capacitive shunt to increase robustness against charge noise at the cost of reduced spectral anharmonicity.
The superconducting flux qubit \cite{orlandoPRB99} offers an alternative that can exhibit large anharmonicity, and researchers continue to refine its design \cite{Yan2016}.   
One innovative reconsideration of the flux qubit, called "fluxonium," was proposed \cite{manucharyanSCI09} to suppress charge-noise sensitivity in all of the eigenstates of the system.  In this paper, we present a potent optimization of the fluxonium design that minimizes its charge-noise decoherence.

\begin{figure}[t]
\begin{center}
	\subfloat[\label{Fig:Cartoon}]{\includegraphics[width=0.25\columnwidth]{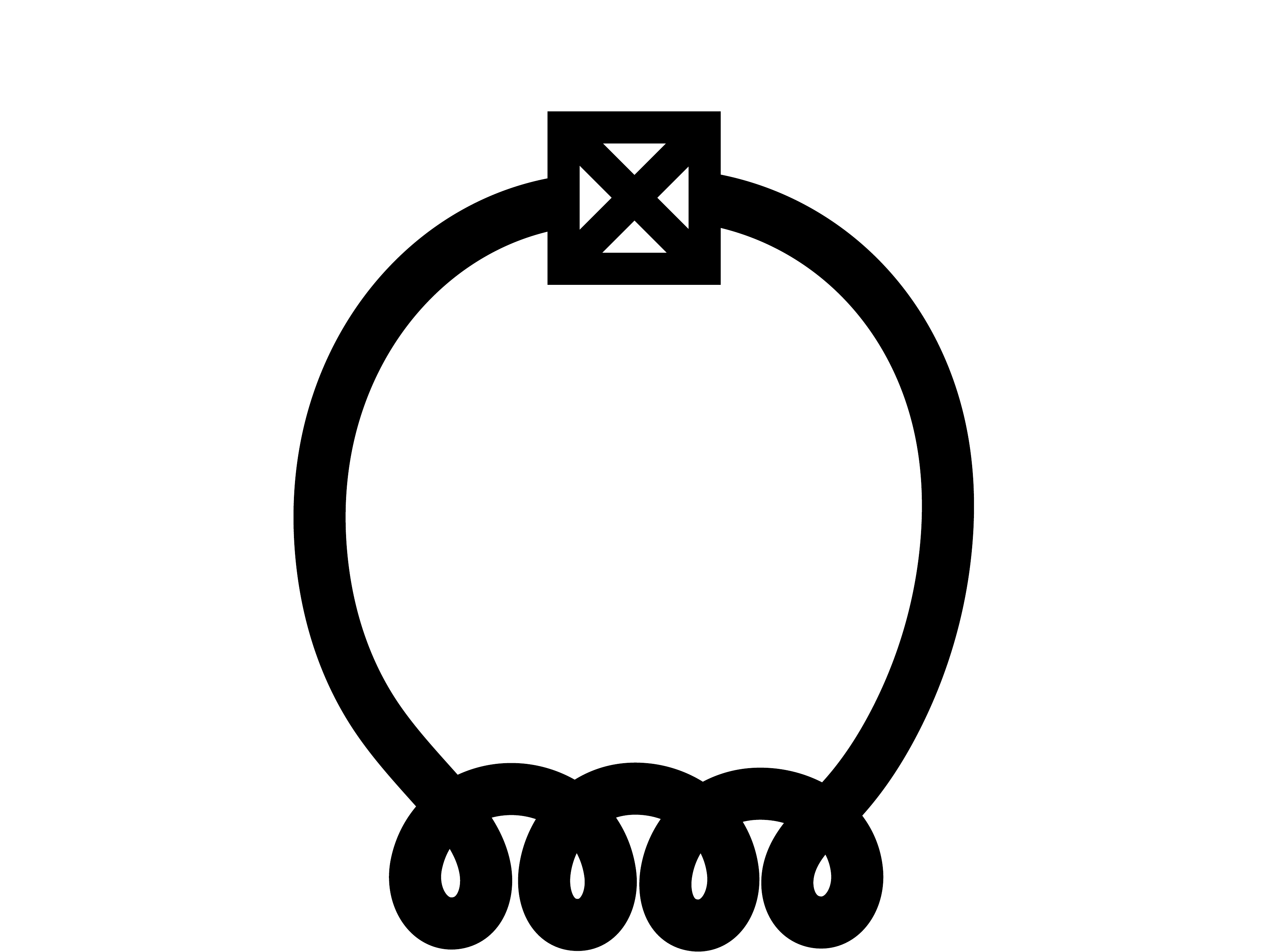}}
	\subfloat[\label{Fig:circuit}]{
		\def\svgwidth{0.75\columnwidth}
		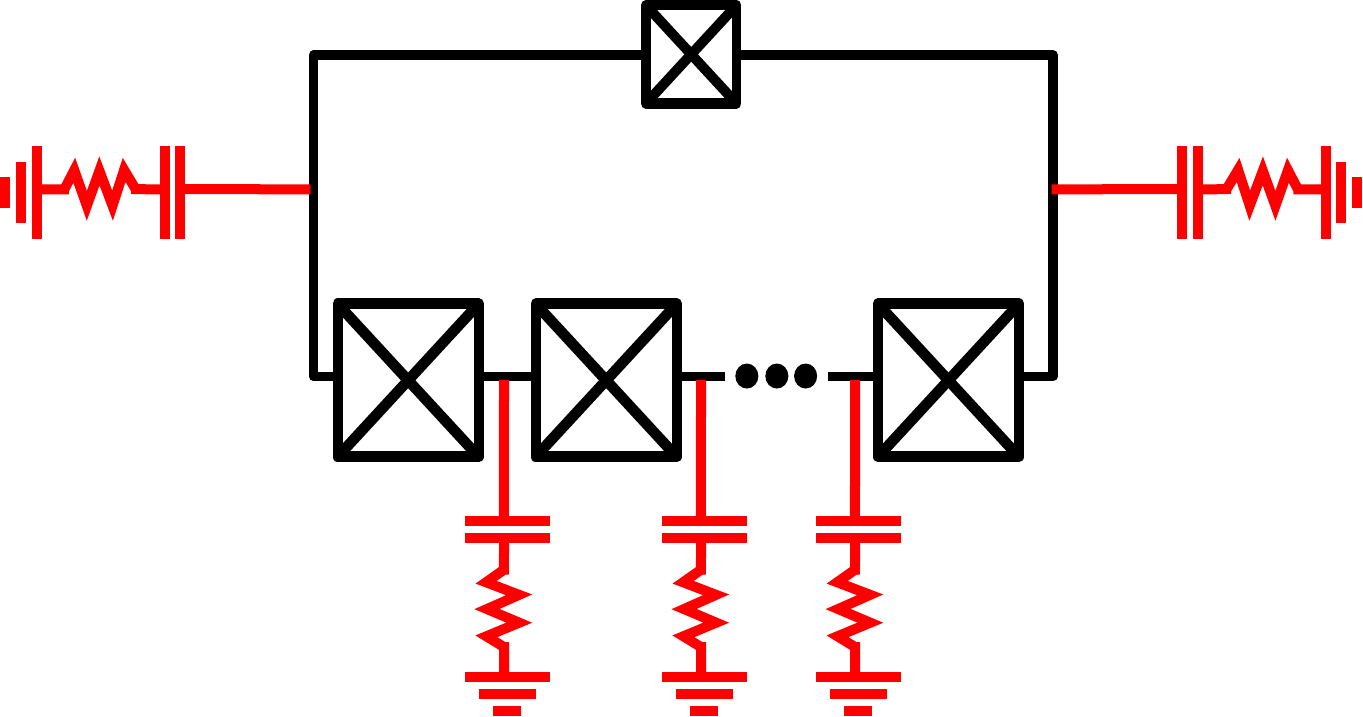
	}
\end{center}
    \caption{The fluxonium qubit. 
    \protect\subref{Fig:Cartoon} Sketch of fluxonium.  Single island of superconductor is interrupted by a single Josephson junction.  The body of the island carries an inductance. 
    \protect\subref{Fig:circuit} Fluxonium circuit diagram.  A ``black-sheep'' junction \cite{ManucharyanPRB2012} with Josephson energy $E_J^b$ and capacitance $C^b$ is shunted by an array of junctions with Josephson energy $E_J^a$ and capacitance $C^a$.  To model dissipation (circuit elements in red), the superconducting array islands and the end islands are coupled to ground via capacitances $C^a_d$ and $C^b_d$ respectively.  An impedance, placed in series with each ground capacitor, causes its voltage drop to fluctuate \cite{Devoret1997}.  This models ambient charge noise.
    }
    \label{Fig:Diagram}
\end{figure}

Fluxonium exploits the fact that a single piece of metal naturally keeps its voltage uniform even in the presence of static external electric fields.  So, instead of a standard flux qubit comprised of three or four superconducting islands, one imagines a qubit constructed from a single, annulus-shaped, island.  The annulus is interrupted by a Josephson junction, and the body of the annulus shunts that junction with a large inductance, as in Fig. \ref{Fig:Cartoon}.  Since it is made of a single island of metal, such a qubit should remain indifferent to low-frequency charge noise.

In practice, the inductance of such a loop is too small to permit a good qubit.  To produce the required large inductance, fluxonium adds a long chain of islands to the loop \cite{manucharyanSCI09}, strongly coupled via Josephson junctions, as in Fig. \ref{Fig:circuit}.  This design choice requires deliberation -- our qubit was motivated by the robustness of a continuous piece of metal, so it seems counterproductive to incorporate a large number of islands.  In the following, we confirm that, as long as the islands are coupled together sufficiently strongly, they can behave like a single piece of superconductor as far as low-frequency charge noise is concerned.  However, using a standard model of charge noise \cite{Yan2016}, we show that the qubit relaxation rate scales with the number of islands.  This leads to our main result: for given fluxonium qubit parameters, there is an optimal number of islands that maximizes the qubit's decoherence time $T_2$.  We focus here on the original fluxonium proposal \cite{manucharyanSCI09} in which the inductor is formed by a chain of coupled superconducting islands, but our findings may be relevant for alternative realizations of the inductor \cite{Hazard2019, Grunhaupt2019,Niepce2019} provided they can be modeled  \cite{matveevPRL02,Maleeva2018} by such a chain.

%

\pardash{Hamiltonian }To calculate the charge-noise decoherence rate of fluxonium, consider the superconducting circuit depicted in Fig. \ref{Fig:circuit}.  The loop is pierced by time-independent flux $\Phi$, so that $\varphi \equiv 2e \Phi/\hbar$ is dimensionless.  We have labelled the gauge-invariant phase drops as shown. The Lagrangian associated with this circuit is $
 \mathcal{L} = \mathcal{T} + \mathcal{T}_d - \mathcal{U} $. Here, the Josephson energy is
\begin{equation} 
\mathcal{ U} = E_J^a \sum_{i=1}^N (1 - \cos \Theta_i)+ E_J^b (1-\cos(\sum_{i=1}^N \Theta_i - \varphi).
\label{Eq:LU}
\end{equation}
The capacitative energy is composed of two parts. The first describes the capacitors around the superconducting loop,
\begin{equation}
\mathcal{T} = \frac{1}{2} C^a \sum_{i=1}^N\left(\frac{\hbar \dot{\Theta}_i}{2e}\right)^2  + \frac{1}{2} C^b \left(\sum_{i=1}^N \frac{\hbar \dot{\Theta}_i}{2e}\right)^2.
\label{Eq:LT}
\end{equation}
The second part, $\mathcal{T}_d$, describes capacitive coupling to dissipative elements.  These dissipative elements, modeled as impedances \cite{Devoret1997,Caldeira1983}, are shown in red in Fig. \ref{Fig:circuit}.  Placed in series with small capacitances to ground $C_d^a$ and $C_d^b$, they produce voltage fluctuations $V_i$ that model background charge noise \cite{Yan2016}.  The associated energy is
\begin{equation} \begin{split}
& {\cal T}_d  = \sum_{i=1}^{N-1} \frac{1}{2} C_d^a \Big(\frac{\hbar \dot{\tau}}{2e} + \sum_{j=1}^{i}\frac{\hbar \dot{\Theta}_j}{2e} - V_i\Big)^2
\\ & + \frac{1}{2} C_d^b \Big(\frac{\hbar \dot{\tau}}{2e}- V_0\Big)^2 
+ \frac{1}{2} C_d^b \Big(\frac{\hbar \dot{\tau}}{2e} + \sum_{j=1}^{N}\frac{\hbar \dot{\Theta}_j}{2e} - V_N\Big)^2.
\label{Eq:LTd}
\end{split}\end{equation}

We define (dimensionless) canonical momenta $\mathcal{N}_\tau = \partial \mathcal{L}/\partial \hbar \dot{\tau}$ and $\mathcal{N}_i = \partial \mathcal{L}/\partial \hbar \dot{\Theta}_i$.  Physically, the $\mathcal{N}_i$ are integer-valued variables determined by the number of Cooper pairs residing on islands of the circuit.  A standard Legendre transformation yields the Hamiltonian 
\begin{equation}
\mathcal{H} = \frac{1}{2}  \left[\begin{array}{c} 2e \mathcal{N}_\tau +  Q_\tau\\ 2e \mathcal{N}_1 + Q_1\\ \vdots \\  2e \mathcal{N}_N + Q_N \end{array} \right]^T {\cal C}^{-1}  \left[\begin{array}{c} 2e \mathcal{N}_\tau +  Q_\tau\\ 2e \mathcal{N}_1 + Q_1\\ \vdots \\  2e \mathcal{N}_N + Q_N \end{array} \right] + {\cal U}
\label{Eq:H}
\end{equation}
where $\mathcal{C}$ is the capacitance matrix obtained from \cref{Eq:LT,Eq:LTd} and the offset charges have the form
\begin{equation} \label{Eq:Qi}
Q_i = C_d^a \sum_{j=i}^{N-1} V_j + C_d^b V_N\,\,\,\,\,\, \text{    and   }\,\,\,\,\,\,\,
Q_\tau = C_d^b V_0 + Q_1.
\end{equation}
The fact that (\ref{Eq:Qi}) involves sums of voltages leads to larger offset charge fluctuations than one might naively assume.  This plays a central role in making the fluxonium relaxation rate increase with $N$ as we show below.  

To analyze $\mathcal{H}$, first note \cite{FergusonPRX2013} that $\mathcal{N}_{\tau}$ is a conserved quantity since $\tau$ is absent from the Josephson energy $\mathcal{U}$.  We can therefore set $\mathcal{N}_\tau$ to zero in $\mathcal{H}$, restricting our attention to eigenstates of $\mathcal{H}$ that are independent of $\tau$. 

Next, we specialize to the case of large $E_J^a$, which is suitable for fluxonium \cite{manucharyanSCI09}.  The low-energy eigenstates then reside in the region $\Theta_{i}\ll 1$, and we approximate $\cos \Theta_{i} \approx 1 - \Theta_{i}^{2}/2$.  This renders the Hamiltonian mostly harmonic.  Following the usual procedure for harmonic Hamiltonians, we introduce a real unitary (orthogonal) transformation $U$ to define $N$ new variables $n_i = \sum_{j=1}^N U_{i j} \mathcal{N}_j$, $q_i = \sum_{j=1}^N U_{i j} Q_j$, and $\theta_i  =\sum_{j=1}^N U_{i j} \Theta_j$.   We set $U_{1 j} = 1/\sqrt{N}$, so that $\theta_1 = \sum_{j=1}^N  \Theta_j/\sqrt{N}$ is an equal superposition mode.  Then, the Hamiltonian decomposes to $\mathcal{H}_{\rm eff} = \sum_{i=1}^{N} H_{i}$, where
\begin{equation}\begin{split}
H_1 &=  \frac{\left(2e n_1 + q_1 - \sqrt{N} Q_\tau/2\right)^2 }{2 (C^a + N C^b)}\\
& + \frac{E_J^a}{2} \theta_1^2 + E_J^b (1 - \cos(\sqrt{N} \theta_1 - \varphi)),
\end{split}\end{equation}
\begin{equation}
H_{i\ne 1} =  4E_{C}^{a} (n_i+\frac{q_i}{2e})^2 + \frac{1}{2}E_{J}^{a} \theta_i^2.
\label{Eq:initialHi}
\end{equation}
Here, we have neglected the effect of $C_d^a$ and $C_d^b$ on the capacitance denominators and have defined $E_{C}^{a} = e^2/2C^a$.

The form of $H_{1}$ becomes more familiar if we set ${\theta = \sqrt{N} \theta_1}$, ${n =  n_1/\sqrt{N}}$, ${q =  q_1/\sqrt{N} -  Q_\tau/2}$.  Then,
\begin{equation} \label{Eq:h}
H_{1} = 4 E_C  \left(n +\frac{q}{2e}\right)^2  + E_J (1 - \cos(\theta - \varphi)) + \frac{E_L \theta^2}{2},
\end{equation}
where $E_C = e^2 / 2(C^a/N + C^b)$, $E_J = E_J^b$, and ${E_L = E_J^a/N}$.  These effective parameters determine the physics of the qubit.  Keeping them fixed, the same Hamiltonian (\ref{Eq:h}) can be realized for different $N$ provided the array junction parameters vary as $C^a = N (e^2/2E_C - C^b)$ and $E_J^a = N E_L $.  Physically, $C^a$ and $E_J^a$ can be tuned this way by simply changing the area of the array junctions.  The central problem we address in this paper is to optimize the charge-noise robustness of fluxonium as a function of $N$.

To make the dependence on $N$ explicit, we rewrite \cref{Eq:initialHi} as
\begin{equation}
H_{i\ne 1} =  \frac{4}{N} \mathcal{E}_C^a (n_i+\frac{q_i}{2e})^2 + \frac{ N E_L }{2} \theta_i^2
\label{Eq:Hi}
\end{equation}
 with $\mathcal{E}_C^a \equiv N E_{C}^{a} = 1/(1/E_C -  2C^b/e^2)$ independent of $N$.  
 
\pardash{Approximate solution}The approximate Hamiltonian $\mathcal{H}_{\text{eff} }$ is conveniently separated in terms of the new variables $\theta, \theta_2,\dots,\theta_N$, so we can find the eigenstates of each term individually.  We denote the Gaussian ground state of the Hamiltonian (\ref{Eq:Hi}) by $e^{-i\theta_{i} q_{i}/2e} \phi_0(\theta_{})$, and the ground state and first excited state of  \cref{Eq:h} by $e^{-i\theta q/2e} \psi_0(\theta)$ and $e^{-i\theta q/2e}\psi_1(\theta)$, respectively.   These states satisfy the usual boundary conditions, vanishing as $\theta \rightarrow \pm \infty$.

 At first, it appears that  the exact ground state of  $\mathcal{H}_{\text{eff} }$ is
\begin{equation}\begin{split}\label{Eq:chi0}
\chi_0 & (\theta,\{\theta_i\})  = e^{-i(q \theta +\sum_{i=2}^{N} q_i \theta_i)/2e}   \psi_0(\theta) \prod_{j=2}^{N}  \phi_0(\theta_j ).
\end{split}\end{equation}
The phase factor in front removes the background charges from $\mathcal{H}_{\text{eff}}$, so that its ground state energy is perfectly independent of low-frequency charge noise.  In fact, this phase factor can be placed in front of every eigenstate of $\mathcal{H}_{\text{eff}}$, so the entire energy spectrum seems to be independent of low-frequency charge noise, the goal described in the second paragraph of this paper.

Upon reflection, we realize that $\chi_0$ unfortunately does not satisfy the correct boundary conditions.  Physically, each superconducting island of the circuit must house an integral number of Cooper pairs.  It follows that $\Theta_i$, since it is conjugate to the discrete variable $\mathcal{N}_i$, must be a compact variable.  In other words, changing the value of $\Theta_i$ by $2 \pi$ does not describe a different state of the system, so the quantum mechanical wavefunctions of the circuit must satisfy periodic boundary conditions in $\Theta_i$.

To address this, we impose the correct boundary conditions using an (unnormalized) tight-binding ansatz \footnote{To carefully verify that  $\Psi_0$ satisfies the correct boundary conditions, regard the new variables $\theta, \theta_2, \dots, \theta_N$ as functions of the original variables $\Theta_i$.  Add $2 \pi$ to any $\Theta_i$, and replace the index $k_i$ everywhere with $k^\prime_i -1= k_i$; $\Psi_0$ returns to itself.},
\begin{equation}\begin{split}
 \label{Eq:Psi0}
&  \Psi_0  (\theta, \{\theta_{i}\})= 
 \\ &  \sum_{\mathclap{k_1= -\infty}}^\infty\;\dotsi\;  \sum_{\mathclap{k_{N}= -\infty}}^\infty
   \chi_0 ( \theta+2 \pi \sum_{j=1}^{N} k_j ,\{\theta_i+ 2 \pi \sum_{j=1}^{N} U_{i,j} k_j \}).
\end{split}\end{equation}
 The $k_1 = \dotsi = k_N=0$ term of $\Psi_0$ is our earlier ground state $\chi_0$.  Since the remaining terms overlap relatively weakly with it (recall $\phi_0$ is strongly localized), $\Psi_0$ is approximately an eigenstate of  $\mathcal{H}_{\text{eff}}$.

We have argued that the sum of terms  in $\Psi_0$ is essential in order to enforce the periodic boundary conditions, without which the spectrum would be independent of the charge-offsets $Q_i$ \cite{KochPRA2007}.  An alternative perspective  is that these terms describe coherent phase-slips in which $\theta$ jumps by a multiple of $2 \pi$.  It is important to stress that these are two ways of looking at the same physical effect:  fluxonium phase-slip physics \cite{ManucharyanPRB2012} is properly incorporated in our analysis.

Note that by substituting $\psi_1$ for $\psi_0$ in \cref{Eq:chi0} to define $\chi_1$, and then $\chi_{1}$ for $\chi_{0}$ in \cref{Eq:Psi0}, we can find the wavefunction of the first excited state, $\Psi_1$. This ansatz for $\Psi_1$ is not perfectly orthogonal with $\Psi_{0}$, but orthogonalizing it leads to negligible corrections.

\pardash{Pure dephasing}
With the approximate wavefunction (\ref{Eq:Psi0}), we can calculate the qubit decoherence time. We first quantify the pure dephasing of the qubit by low-frequency charge noise. Dephasing occurs when a shift in the charge parameters $Q_i$ alters  the transition frequency of the qubit, ${\omega_{01} = (E_1-E_0)/\hbar}$.

To find the dependence $\omega_{01}$ on the offset charges, we calculate the expectation of the original, periodic Hamiltonian $\mathcal{H}$ in state $\Psi_n$ and find that it varies with $Q_i$ as
\begin{align}  \nonumber
& E_{n}(Q_\tau,\dots,Q_N) = \\
& E_{n}(0,\frac{e}{2},\dots,\frac{e}{2}) - \frac{\epsilon_n}{2 } \sum_{j=1}^N \cos \frac{2 \pi Q_j - \pi Q_\tau}{2e},
\label{Eq:E0}
\end{align}
for $n=0,1$, where
\begin{widetext}
\begin{equation}\begin{split}
\epsilon_n=  4 N E_L  e^{-\pi^2 \sqrt{E_L/8{\cal E}_{C}^{a}} (N-1)} & \int_{-\infty}^{\infty} d\theta \psi_n^{*}(\theta + 2 \pi) \psi_n(\theta) 
	\left[  \frac{\pi^2}{2} \left( 1 - \frac{1}{N}\right)  + \left(\frac{\theta^2}{2N} - \int_{-\infty}^{\infty} d \bar{\theta} \left| \psi_n(\bar{\theta}) \right|^2 \frac{\bar{\theta}^2}{2N}\right)
	 \right. \\ \label{Eq:epsilon0}
& \left. \qquad + e^{-\sqrt{{\cal E}_{C}^{a}/2E_L} (N-1)/N^2} \left((N-2) \cos  \frac{\theta + \pi}{N}  - \int_{-\infty}^{\infty} d \bar{\theta} \left| \psi_n(\bar{\theta}) \right|^2 N \cos \frac{\bar{\theta}}{N}\right)
  \right] .
\end{split}\end{equation}
\end{widetext}
This expression is derived within the tight-binding approximation, neglecting matrix elements between next nearest neighbor terms of \cref{Eq:Psi0} and beyond.  
 
We can now find the qubit's dephasing rate. For simplicity, suppose all of the voltages $V_i$ have the same noise power spectrum $S_V(\omega)$, and let the associated charge fluctuation be 
\begin{equation}
S_{\text{charge}}(\omega) \equiv (C_d^a)^2 S_{V}(\omega).
\label{Eq:S}
\end{equation}
Assume a $1/f$ form for the low-frequency power spectrum $S_{\text{charge}}(\omega) = 2 \pi A^2_{\text{charge}}/| \omega|$.   Then, reasoning as in \cite{KochPRA2007}, we find 
\begin{equation}
\frac{1}{T_\phi} \sim \sqrt{\frac{N}{2}\left(\frac{N-1}{2} + \left(\frac{C_d^b}{C_d^a}\right)^2 \right)} \frac{\left| \epsilon_1 - \epsilon_0\right|}{\hbar}\frac{A _{\text{charge}} \pi}{2 e}.
\label{Eq:Tphi}
\end{equation}
The pure dephasing time rapidly increases with $N$ because of the exponential factor in \cref{Eq:epsilon0}, as predicted in \cite{manucharyanSCI09}.  This is because the ratio $E_J^a /E_{C}^{a} = N^{2} E_L/{\cal E}_C^a$ increases with $N$ for fixed $E_L$ and ${\cal E}_{C}^{a}$, carrying the superconducting islands further into the transmon regime \cite{KochPRA2007}.  Alternatively, the increasing value of $E_J^a$ means stronger coupling between superconducting islands, which therefore better approximate the single piece of metal discussed in the second paragraph of this paper.

\pardash{Relaxation rate}  The other source of decoherence is unwanted transitions between the two computational states. To compute the rate of this relaxation, consider the term $4 E_C n q /e$ obtained by expanding the square in  \cref{Eq:h}.  The qubit lifetime is determined by the matrix element of this term between $\Psi_0$ and $\Psi_1$.   The offset charge $q$ varies with the fluctuating voltages $V_i$ \footnote{Note $q = C_d^a \sum_{k=1}^{N-1} (k/N-1/2) V_k + C_d^b(V_N-V_0)/2$.  This is unchanged, as to be expected, by a constant shift of all $V_i$.}, leading to 
\begin{equation}\begin{split}
\frac{1}{T_1} &= \frac{8E_C^2}{\hbar^2} \left| \left<\psi_0\right|n\left|\psi_1\right>\right|^2 \times\\
& \left( \frac{(N-2)(N-1)}{6N}+ \left(\frac{C_d^b}{C_d^a}\right)^2\right) \frac{S_{\text{charge}}(\omega_{01})}{e^2}.
\label{Eq:T1}
\end{split}\end{equation}
within the tight-binding approximation. In contrast to the dephasing time, we discover that $T_1$ \emph{decreases} with $N$.

\pardash{Net decoherence rate} We incorporate \cref{Eq:Tphi,Eq:T1} into the net decoherence rate using the standard relation $1/T_2 = 1/T_\phi + 1/2T_1$.  Since the relaxation rate increase with the number of voltages while the pure dephasing rate decreases, $T_2$ has a maximum with respect to $N$.   

\begin{figure}[ht]
\begin{center}
\includegraphics[width=2.75in]{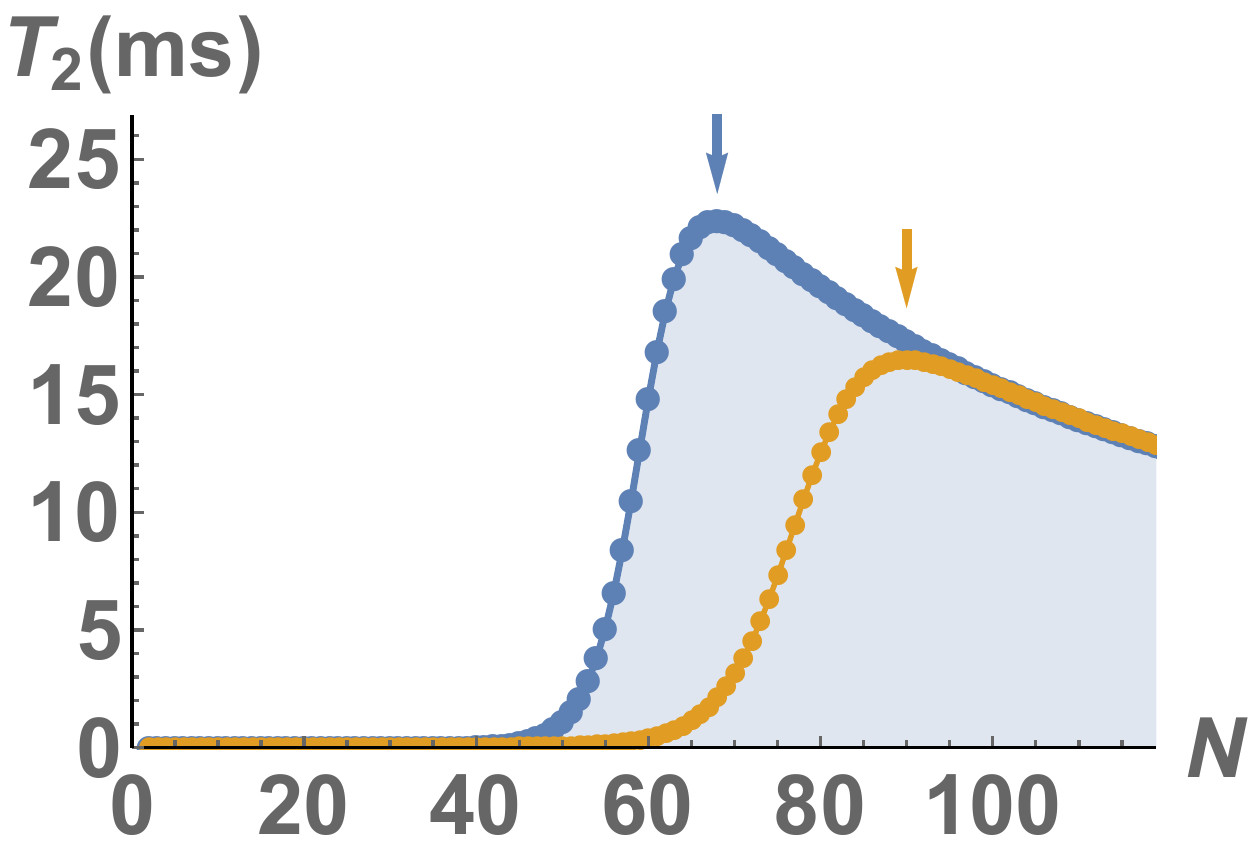}
\end{center}
    \caption{Plot of $T_2$ versus number of islands $N$.  Fluxonium parameters are fixed at $E_C = 2.5$, $E_J = 9.0$, and $E_L = 0.52$ GHz, as in experiment \cite{manucharyanSCI09}.  These imply $e^2/2C^b = 2.64$ GHz.  Blue arrow indicates optimal choice $N=68$.
    The yellow curve shows a recalculation to check the effect of reduced wavefunction confinement, see \cref{Eq:broadphi}, with the arrow indicating optimal choice $N=90$.}
    \label{Fig:T2versusNoriginal}
\end{figure}

\begin{figure}[ht]
\begin{center}
\includegraphics[width=2.75in]{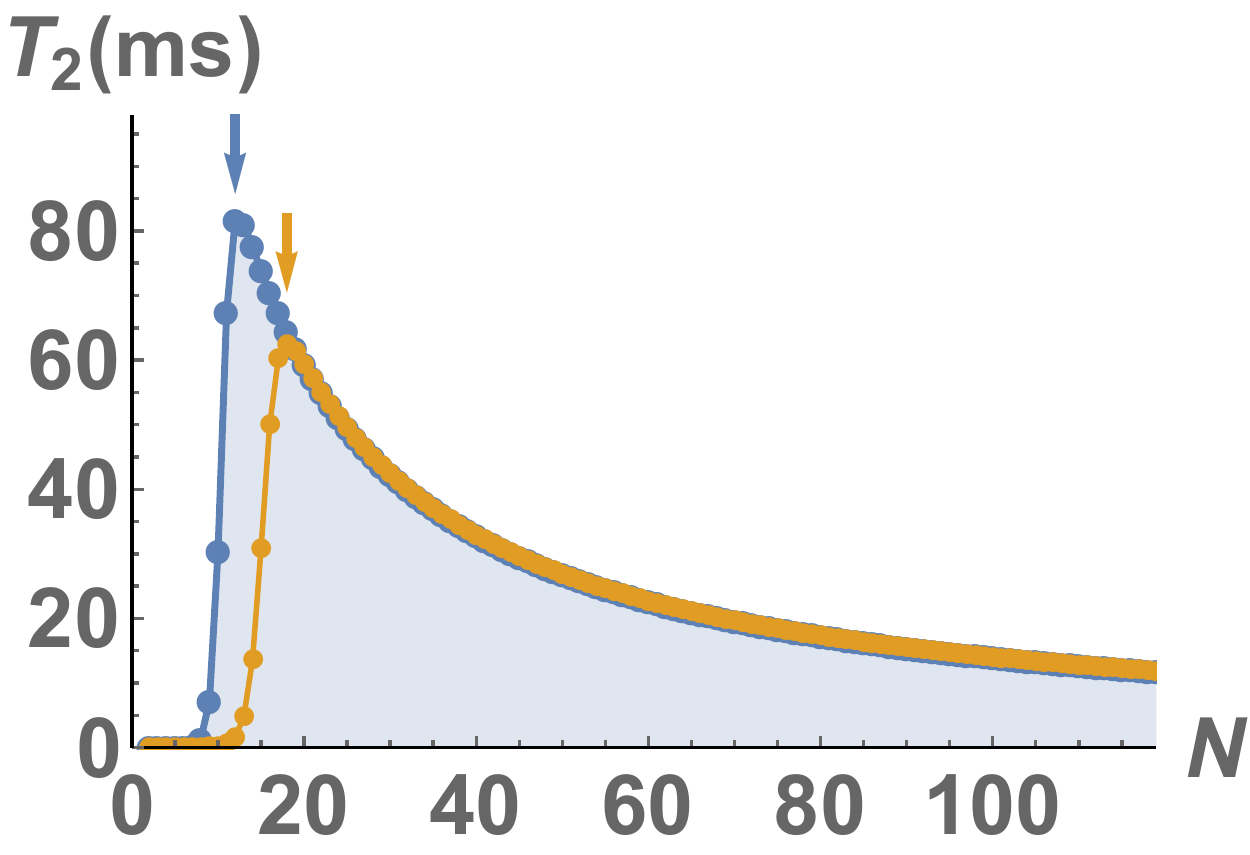}
\end{center}
    \caption{Plot of $T_2$ versus number of islands $N$.  Fluxonium parameters are fixed at $E_C = 0.55$, $E_J =2.2$, and $E_L =0.72$ GHz, as in device C of experiment \cite{Nguyen2018}.   These imply $e^2/2C^b = 0.73$ GHz.  Blue arrow indicates optimal choice $N=12$.
	The yellow curve shows a recalculation to check the effect of reduced wavefunction confinement, see \cref{Eq:broadphi}, with an arrow indicating optimal choice $N=18$.  Similar results are obtained for the other devices in \cite{Nguyen2018}.}
    \label{Fig:T2versusNdeviceC}
\end{figure}

In Figs. \ref{Fig:T2versusNoriginal} and \ref{Fig:T2versusNdeviceC}, we show the dependence of $T_2$ upon $N$ using values of $E_C$, $E_J$, and $E_L$, and $C^b$ from fluxonium experiments \cite{manucharyanSCI09,Nguyen2018}.  We have chosen ${C^b= (e^2/2E_C)/(1 + E_L/E_J)}$, independent of $N$.  This follows from ${C^b/C^a = E_J^b/E_J^a =  E_J/N E_L}$, which is true since capacitance and Josephson energy both scale with junction area.  The flux through the loop is set to  $\varphi= \pi$.  For the low-frequency noise spectrum in \cref{Eq:Tphi}, we set $A_{\rm charge} = 10^{-3} e$ \cite{Zorin1996,KochPRA2007,Krantz2019}.  For the high-frequency power spectrum in \cref{Eq:T1}, we adopt the ohmic charge noise model \cite{Yan2016}  $S_{\text{charge}}(\omega) = {\cal A}_{\text{charge}}^2 \omega/(2 \pi \times  \text{1 GHz})$, with ${{\cal A}_{\text{charge}}^2 = (5.2 \times 10^{-9} e)^2/\text{Hz}}$.  For simplicity, we set $C^b_d = C^a_d$.  The rates (\ref{Eq:Tphi}) and (\ref{Eq:T1}) are evaluated using numerically computed fluxonium wavefunctions $\psi_i$.

Fig. \ref{Fig:T2versusNoriginal} considers the early fluxonium experiment \cite{manucharyanSCI09}.  The blue curve indicates that, for the experimentally chosen parameters of $E_C$, $E_J$, $E_L$, and $C^b$, the optimal value of $N$ is $68$. The original device, with $N=43$, had a ratio of only $E_J^a/E_{C}^{a}\approx 22$, leading to excessive low-frequency charge noise dephasing.  At the optimal $N$, $E_J^a/E_{C}^{a} \approx 53$, so the array junctions are deeper in the  transmon regime, leading to better suppression of charge-noise dephasing.

Fig. \ref{Fig:T2versusNdeviceC} provides an analogous plot for a recent experiment \cite{Nguyen2018}.  The blue curve shows that the optimal choice is the relatively small value $N=12$, far less than the experimental value $N=102$.  This striking reduction arises since the original device had array junctions with $E_J^a/E_{C}^{a}>3000$, far larger than needed to protect against low-frequency charge noise. The satisfactory value $E_J^a/E_{C}^{a}\approx 47$ is achieved at $N=12$; further increasing $N$ just brings about a faster relaxation rate $1/T_1$.

\begin{figure}[ht]
\begin{center}
\includegraphics[width=2.75in]{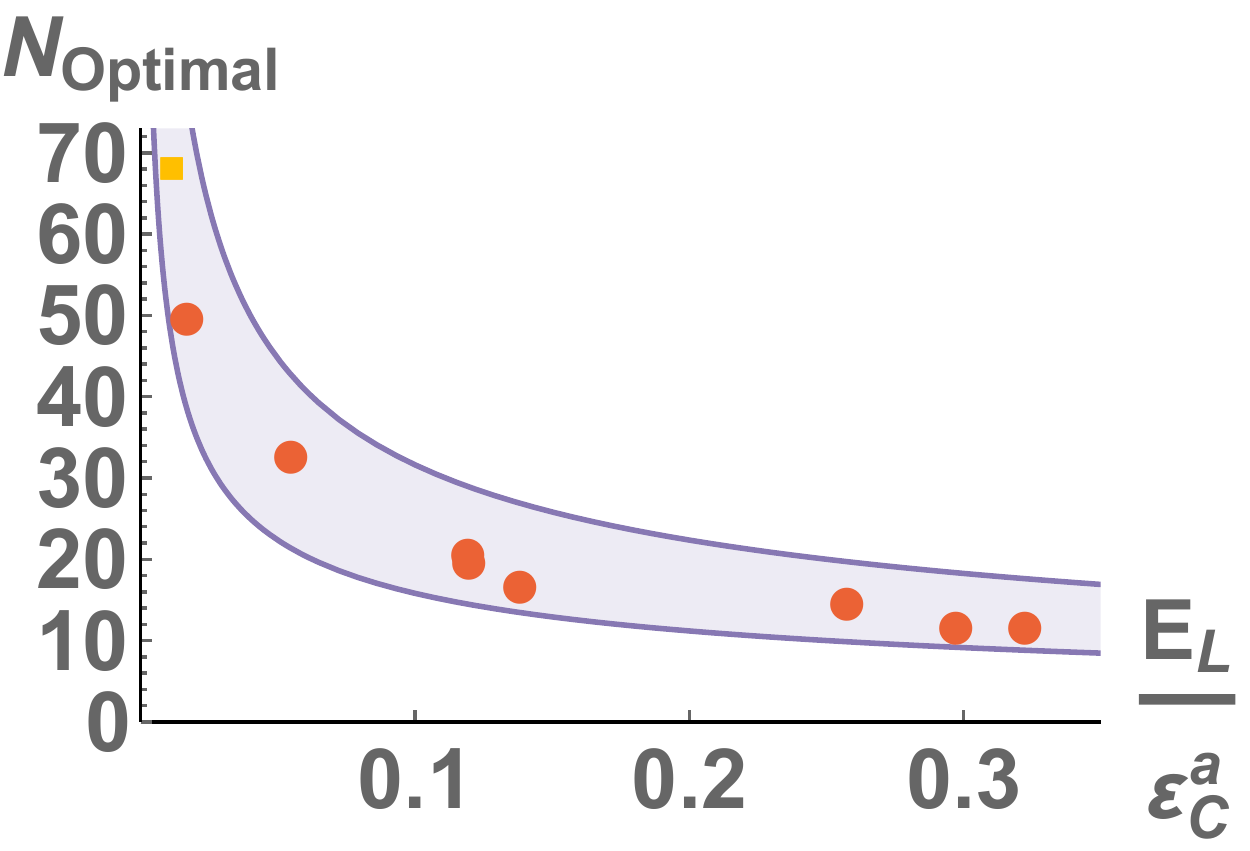}
\end{center}
    \caption{The optimal number of junctions $N$ as a function of the ratio $E_L/{\cal E}_C^a$.  The square yellow data point on the left is taken from \cite{manucharyanSCI09} while the remaining 8 data points are taken from \cite{Nguyen2018}.  The region between the curves $5/\sqrt{E_L/{\cal E}_C^a}$ and $10/\sqrt{E_L/{\cal E}_C^a}$ is shaded, showing agreement with the rule-of-thumb for $N_{\text{optimal}}$.}
    \label{Fig:optimumNversusratio}
\end{figure}

This kind of argument gives a general rule-of-thumb for the optimal $N$.  Because the pure dephasing rate in \cref{Eq:epsilon0,Eq:Tphi} drops exponentially with $N$ while the relaxation rate in \cref{Eq:T1} increases only polynomially, the optimal number of junctions is just large enough to suppress the former. As shown in \cref{Fig:optimumNversusratio}, this means that up to a logarithmic correction, $N_{\text{optimal}} \sim 5-10\times 1/\sqrt{E_{L}/{\cal E}_{C}^{a}}$. This value ensures that the array junctions are sufficiently ``transmon-like'' with $E_J^a/E_{C}^{a} \gtrsim 50$.  The optimal fluxonium qubit incorporates the minimal number of junctions consistent with this constraint and the desired $E_C$, $E_J$, and $E_L$.

Naturally, the millisecond-scale $T_2$ times in Figs. \ref{Fig:T2versusNoriginal} and \ref{Fig:T2versusNdeviceC} exceed the much shorter values measured experimentally.  This is unsurprising, because the figures only consider charge noise, neglecting all other mechanisms of decoherence.  One expects that such mechanisms, like flux noise or Purcell emission, are functions of $E_C$, $E_L$, and $E_J$ that probably do not depend sensitively on $N$.  It is plausible that some mechanisms could favor smaller fluxonium designs, in which case the optimal value $N = 12$ found in Fig. \ref{Fig:T2versusNdeviceC} could only decrease.

\pardash{Discussion }

The optimizations here should not require experimentally unrealistic parameters.  We have fixed $C^b$, $E_J^b$ near their experimental values  \cite{manucharyanSCI09,Nguyen2018} while the values required for $C^a$ and $E_J^a$ should be attainable by scaling the area of the array junctions.  For example, to realize the optimal point $N=12$ in Fig. \ref{Fig:T2versusNdeviceC}, the array junctions should be modified to have an area about $12/102$ times what was chosen in the $N=102$ experiment.  The resulting junctions would still be larger than the ``black-sheep'' junction (i.e., at the optimum $N$, $C^a > C^b$ and $E_J^a > E_J^b$).   In any case, the superconducting qubit platform is characterized by remarkable experimental flexibility.  There are many possibilities one could imagine to realize specified junction parameters, such as shunting each of the array junctions with its own transmon-style capacitor  \cite{KochPRA2007}.

One expects generally correct answers from the harmonic approximation (\ref{Eq:initialHi}) and the tight-binding approximation (\ref{Eq:Psi0}) that underlie our calculations.   Symmetry considerations decidedly limit the effect of corrections to the harmonic approximation, as investigated thoroughly in \cite{FergusonPRX2013}.  In addition, as discussed above, the optimal $N$ depends only logarithmically on most of the parameters in the equation.  
However, some caution is appropriate --  the harmonic approximation does exaggerate the confining potential, since  $1-\cos \Theta_i \le \Theta_i^2/2$.  The Gaussian ground state wavefunctions $\phi_0$ are thus overly localized, suppressing the overlap between neighboring terms in \cref{Eq:Psi0}.  As a result, the matrix element (\ref{Eq:epsilon0}) and dephasing rate (\ref{Eq:Tphi}) are somewhat underestimated.  

To assess the amount of error that results, we consider the broadened Gaussian wavefunctions
\begin{equation}
\phi_0(\theta_i)\to \sqrt{\lambda} \phi_0(\lambda\theta_i),
\label{Eq:broadphi}
\end{equation}
with $\lambda = 2/\pi$.  These would be the eigenstates of \cref{Eq:initialHi} if we changed $\Theta_{i}^{2}/2 \to 2\Theta_{i}^{2}/\pi^{2}\le 1- \cos\Theta_{i}$ to bound the Josephson potential from below.
We use these revised $\phi_0$ to recalculate \cref{Eq:epsilon0} \footnote{In each of the exponentials of \cref{Eq:epsilon0}, but not the prefactor, $E_L$ gets multiplied by $\lambda^2$.  The term $\frac{\pi^2}{2} \left( 1 - \frac{1}{N}\right)$ also gets multiplied by $\lambda^2$.}.  The result, shown as yellow curves in \cref{Fig:T2versusNoriginal,Fig:T2versusNdeviceC}, is an increase of the optimum $N$ by around ${\pi/2 - 1\approx 50\%}$ and a modest decrease in the associated $T_2$.  
Of course, the yellow curves significantly overestimate the dephasing rate, and we expect the real value to be closer to the blue curves \footnote{It is possible to treat $\lambda$ as a variational parameter in \cref{Eq:Psi0} and minimize the energy of $\mathcal{H}$ to determine $\lambda$ at each $N$.  The obtained $\lambda$ is  generally quite close to 1, leading to curves that hug the blue curves in Figs.  \ref{Fig:T2versusNoriginal} and  \ref{Fig:T2versusNdeviceC}}.  It is also instructive to note that the transmon case, solved exactly using a Mathieu function \cite{KochPRA2007} without a tight-binding approximation, exhibits $\epsilon_{0}\sim e^{-\sqrt{8 E_L /\mathcal{E}_C^a }}$, implying $\lambda \sim 8/\pi^2  \approx 0.8$. This would increase our optimal values of $N$ in Figs.  \ref{Fig:T2versusNoriginal} and  \ref{Fig:T2versusNdeviceC} by around $20\%$, intermediate between the blue and yellow curves.  Taken together, these checks show that our approximations give the correct picture and only lead to modest quantitative errors.

The findings presented here identify an important potential optimization of the fluxonium design.  Our prediction follows from the familiar charge-noise model  described in Fig. \ref{Fig:circuit}  and is absent from earlier studies of fluxonium decoherence \cite{ManucharyanPRB2012,Viola2015} that considered different forms of environmental noise.  For instance, \cite{ManucharyanPRB2012} instead assumed an admitttance in parallel with each of the Josephson junctions of the circuit, which naturally models dissipative current fluctuations (such as quasiparticle tunneling \cite{Catelani2011,Yan2016}) across the junctions.  Thus, an experimental test of our results could shed light on the charge-noise model of superconducting qubits.  Most importantly, a significant gain in device performance could result from our proposed optimization.

\pardash{Acknowledgements} We are grateful to Ben Palmer for generously sharing his expertise.
\bibliographystyle{apsrev4-1}
\bibliography{../bibliography/physics}

\begin{thebibliography}{26}%
\makeatletter
\providecommand \@ifxundefined [1]{%
 \@ifx{#1\undefined}
}%
\providecommand \@ifnum [1]{%
 \ifnum #1\expandafter \@firstoftwo
 \else \expandafter \@secondoftwo
 \fi
}%
\providecommand \@ifx [1]{%
 \ifx #1\expandafter \@firstoftwo
 \else \expandafter \@secondoftwo
 \fi
}%
\providecommand \natexlab [1]{#1}%
\providecommand \enquote  [1]{``#1''}%
\providecommand \bibnamefont  [1]{#1}%
\providecommand \bibfnamefont [1]{#1}%
\providecommand \citenamefont [1]{#1}%
\providecommand \href@noop [0]{\@secondoftwo}%
\providecommand \href [0]{\begingroup \@sanitize@url \@href}%
\providecommand \@href[1]{\@@startlink{#1}\@@href}%
\providecommand \@@href[1]{\endgroup#1\@@endlink}%
\providecommand \@sanitize@url [0]{\catcode `\\12\catcode `\$12\catcode
  `\&12\catcode `\#12\catcode `\^12\catcode `\_12\catcode `\%12\relax}%
\providecommand \@@startlink[1]{}%
\providecommand \@@endlink[0]{}%
\providecommand \url  [0]{\begingroup\@sanitize@url \@url }%
\providecommand \@url [1]{\endgroup\@href {#1}{\urlprefix }}%
\providecommand \urlprefix  [0]{URL }%
\providecommand \Eprint [0]{\href }%
\providecommand \doibase [0]{http://dx.doi.org/}%
\providecommand \selectlanguage [0]{\@gobble}%
\providecommand \bibinfo  [0]{\@secondoftwo}%
\providecommand \bibfield  [0]{\@secondoftwo}%
\providecommand \translation [1]{[#1]}%
\providecommand \BibitemOpen [0]{}%
\providecommand \bibitemStop [0]{}%
\providecommand \bibitemNoStop [0]{.\EOS\space}%
\providecommand \EOS [0]{\spacefactor3000\relax}%
\providecommand \BibitemShut  [1]{\csname bibitem#1\endcsname}%
\let\auto@bib@innerbib\@empty
\bibitem [{\citenamefont {Shnirman}\ \emph {et~al.}(1997)\citenamefont
  {Shnirman}, \citenamefont {Sch\"on},\ and\ \citenamefont
  {Hermon}}]{Schnirman97}%
  \BibitemOpen
  \bibfield  {author} {\bibinfo {author} {\bibfnamefont {A.}~\bibnamefont
  {Shnirman}}, \bibinfo {author} {\bibfnamefont {G.}~\bibnamefont {Sch\"on}}, \
  and\ \bibinfo {author} {\bibfnamefont {Z.}~\bibnamefont {Hermon}},\ }\href
  {\doibase 10.1103/PhysRevLett.79.2371} {\bibfield  {journal} {\bibinfo
  {journal} {Phys. Rev. Lett.}\ }\textbf {\bibinfo {volume} {79}},\ \bibinfo
  {pages} {2371} (\bibinfo {year} {1997})}\BibitemShut {NoStop}%
\bibitem [{\citenamefont {Bouchiat}\ \emph {et~al.}(1998)\citenamefont
  {Bouchiat}, \citenamefont {Vion}, \citenamefont {Joyez}, \citenamefont
  {Esteve},\ and\ \citenamefont {Devoret}}]{Bouchiat_1998}%
  \BibitemOpen
  \bibfield  {author} {\bibinfo {author} {\bibfnamefont {V.}~\bibnamefont
  {Bouchiat}}, \bibinfo {author} {\bibfnamefont {D.}~\bibnamefont {Vion}},
  \bibinfo {author} {\bibfnamefont {P.}~\bibnamefont {Joyez}}, \bibinfo
  {author} {\bibfnamefont {D.}~\bibnamefont {Esteve}}, \ and\ \bibinfo {author}
  {\bibfnamefont {M.~H.}\ \bibnamefont {Devoret}},\ }\href {\doibase
  10.1238/physica.topical.076a00165} {\bibfield  {journal} {\bibinfo  {journal}
  {Physica Scripta}\ }\textbf {\bibinfo {volume} {T76}},\ \bibinfo {pages}
  {165} (\bibinfo {year} {1998})}\BibitemShut {NoStop}%
\bibitem [{\citenamefont {Nakamura}\ \emph {et~al.}(1999)\citenamefont
  {Nakamura}, \citenamefont {Pashkin},\ and\ \citenamefont
  {Tsai}}]{NakamuraNat99}%
  \BibitemOpen
  \bibfield  {author} {\bibinfo {author} {\bibfnamefont {Y.}~\bibnamefont
  {Nakamura}}, \bibinfo {author} {\bibfnamefont {Y.~A.}\ \bibnamefont
  {Pashkin}}, \ and\ \bibinfo {author} {\bibfnamefont {J.~S.}\ \bibnamefont
  {Tsai}},\ }\href {http://dx.doi.org/10.1038/19718} {\bibfield  {journal}
  {\bibinfo  {journal} {Nature}\ }\textbf {\bibinfo {volume} {398}},\ \bibinfo
  {pages} {786} (\bibinfo {year} {1999})}\BibitemShut {NoStop}%
\bibitem [{\citenamefont {Nakamura}\ \emph {et~al.}(2002)\citenamefont
  {Nakamura}, \citenamefont {Pashkin}, \citenamefont {Yamamoto},\ and\
  \citenamefont {Tsai}}]{NakamuraPRL02}%
  \BibitemOpen
  \bibfield  {author} {\bibinfo {author} {\bibfnamefont {Y.}~\bibnamefont
  {Nakamura}}, \bibinfo {author} {\bibfnamefont {Y.~A.}\ \bibnamefont
  {Pashkin}}, \bibinfo {author} {\bibfnamefont {T.}~\bibnamefont {Yamamoto}}, \
  and\ \bibinfo {author} {\bibfnamefont {J.~S.}\ \bibnamefont {Tsai}},\ }\href
  {\doibase 10.1103/PhysRevLett.88.047901} {\bibfield  {journal} {\bibinfo
  {journal} {Phys. Rev. Lett.}\ }\textbf {\bibinfo {volume} {88}},\ \bibinfo
  {pages} {047901} (\bibinfo {year} {2002})}\BibitemShut {NoStop}%
\bibitem [{\citenamefont {Koch}\ \emph {et~al.}(2007)\citenamefont {Koch},
  \citenamefont {Yu}, \citenamefont {Gambetta}, \citenamefont {Houck},
  \citenamefont {Schuster}, \citenamefont {Majer}, \citenamefont {Blais},
  \citenamefont {Devoret}, \citenamefont {Girvin},\ and\ \citenamefont
  {Schoelkopf}}]{KochPRA2007}%
  \BibitemOpen
  \bibfield  {author} {\bibinfo {author} {\bibfnamefont {J.}~\bibnamefont
  {Koch}}, \bibinfo {author} {\bibfnamefont {T.~M.}\ \bibnamefont {Yu}},
  \bibinfo {author} {\bibfnamefont {J.}~\bibnamefont {Gambetta}}, \bibinfo
  {author} {\bibfnamefont {A.~A.}\ \bibnamefont {Houck}}, \bibinfo {author}
  {\bibfnamefont {D.~I.}\ \bibnamefont {Schuster}}, \bibinfo {author}
  {\bibfnamefont {J.}~\bibnamefont {Majer}}, \bibinfo {author} {\bibfnamefont
  {A.}~\bibnamefont {Blais}}, \bibinfo {author} {\bibfnamefont {M.~H.}\
  \bibnamefont {Devoret}}, \bibinfo {author} {\bibfnamefont {S.~M.}\
  \bibnamefont {Girvin}}, \ and\ \bibinfo {author} {\bibfnamefont {R.~J.}\
  \bibnamefont {Schoelkopf}},\ }\href {\doibase 10.1103/PhysRevA.76.042319}
  {\bibfield  {journal} {\bibinfo  {journal} {Phys. Rev. A}\ }\textbf {\bibinfo
  {volume} {76}},\ \bibinfo {pages} {042319} (\bibinfo {year}
  {2007})}\BibitemShut {NoStop}%
\bibitem [{\citenamefont {Orlando}\ \emph {et~al.}(1999)\citenamefont
  {Orlando}, \citenamefont {Mooij}, \citenamefont {Tian}, \citenamefont
  {van~der Wal}, \citenamefont {Levitov}, \citenamefont {Lloyd},\ and\
  \citenamefont {Mazo}}]{orlandoPRB99}%
  \BibitemOpen
  \bibfield  {author} {\bibinfo {author} {\bibfnamefont {T.~P.}\ \bibnamefont
  {Orlando}}, \bibinfo {author} {\bibfnamefont {J.~E.}\ \bibnamefont {Mooij}},
  \bibinfo {author} {\bibfnamefont {L.}~\bibnamefont {Tian}}, \bibinfo {author}
  {\bibfnamefont {C.~H.}\ \bibnamefont {van~der Wal}}, \bibinfo {author}
  {\bibfnamefont {L.~S.}\ \bibnamefont {Levitov}}, \bibinfo {author}
  {\bibfnamefont {S.}~\bibnamefont {Lloyd}}, \ and\ \bibinfo {author}
  {\bibfnamefont {J.~J.}\ \bibnamefont {Mazo}},\ }\href {\doibase
  10.1103/PhysRevB.60.15398} {\bibfield  {journal} {\bibinfo  {journal} {Phys.
  Rev. B}\ }\textbf {\bibinfo {volume} {60}},\ \bibinfo {pages} {15398}
  (\bibinfo {year} {1999})}\BibitemShut {NoStop}%
\bibitem [{\citenamefont {Yan}\ \emph {et~al.}(2016)\citenamefont {Yan},
  \citenamefont {Gustavsson}, \citenamefont {Kamal}, \citenamefont {Birenbaum},
  \citenamefont {Sears}, \citenamefont {Hover}, \citenamefont {Gudmundsen},
  \citenamefont {Rosenberg}, \citenamefont {Samach}, \citenamefont {Weber},
  \citenamefont {Yoder}, \citenamefont {Orlando}, \citenamefont {Clarke},
  \citenamefont {Kerman},\ and\ \citenamefont {Oliver}}]{Yan2016}%
  \BibitemOpen
  \bibfield  {author} {\bibinfo {author} {\bibfnamefont {F.}~\bibnamefont
  {Yan}}, \bibinfo {author} {\bibfnamefont {S.}~\bibnamefont {Gustavsson}},
  \bibinfo {author} {\bibfnamefont {A.}~\bibnamefont {Kamal}}, \bibinfo
  {author} {\bibfnamefont {J.}~\bibnamefont {Birenbaum}}, \bibinfo {author}
  {\bibfnamefont {A.~P.}\ \bibnamefont {Sears}}, \bibinfo {author}
  {\bibfnamefont {D.}~\bibnamefont {Hover}}, \bibinfo {author} {\bibfnamefont
  {T.~J.}\ \bibnamefont {Gudmundsen}}, \bibinfo {author} {\bibfnamefont
  {D.}~\bibnamefont {Rosenberg}}, \bibinfo {author} {\bibfnamefont
  {G.}~\bibnamefont {Samach}}, \bibinfo {author} {\bibfnamefont
  {S.}~\bibnamefont {Weber}}, \bibinfo {author} {\bibfnamefont {J.~L.}\
  \bibnamefont {Yoder}}, \bibinfo {author} {\bibfnamefont {T.~P.}\ \bibnamefont
  {Orlando}}, \bibinfo {author} {\bibfnamefont {J.}~\bibnamefont {Clarke}},
  \bibinfo {author} {\bibfnamefont {A.~J.}\ \bibnamefont {Kerman}}, \ and\
  \bibinfo {author} {\bibfnamefont {W.~D.}\ \bibnamefont {Oliver}},\ }\href
  {https://doi.org/10.1038/ncomms12964} {\bibfield  {journal} {\bibinfo
  {journal} {Nature Communications}\ }\textbf {\bibinfo {volume} {7}},\
  \bibinfo {pages} {12964} (\bibinfo {year} {2016})}\BibitemShut {NoStop}%
\bibitem [{\citenamefont {Manucharyan}\ \emph {et~al.}(2009)\citenamefont
  {Manucharyan}, \citenamefont {Koch}, \citenamefont {Glazman},\ and\
  \citenamefont {Devoret}}]{manucharyanSCI09}%
  \BibitemOpen
  \bibfield  {author} {\bibinfo {author} {\bibfnamefont {V.~E.}\ \bibnamefont
  {Manucharyan}}, \bibinfo {author} {\bibfnamefont {J.}~\bibnamefont {Koch}},
  \bibinfo {author} {\bibfnamefont {L.~I.}\ \bibnamefont {Glazman}}, \ and\
  \bibinfo {author} {\bibfnamefont {M.~H.}\ \bibnamefont {Devoret}},\ }\href
  {\doibase 10.1126/science.1175552} {\bibfield  {journal} {\bibinfo  {journal}
  {Science}\ }\textbf {\bibinfo {volume} {326}},\ \bibinfo {pages} {113}
  (\bibinfo {year} {2009})}\BibitemShut {NoStop}%
\bibitem [{\citenamefont {Hazard}\ \emph {et~al.}(2019)\citenamefont {Hazard},
  \citenamefont {Gyenis}, \citenamefont {Di~Paolo}, \citenamefont {Asfaw},
  \citenamefont {Lyon}, \citenamefont {Blais},\ and\ \citenamefont
  {Houck}}]{Hazard2019}%
  \BibitemOpen
  \bibfield  {author} {\bibinfo {author} {\bibfnamefont {T.~M.}\ \bibnamefont
  {Hazard}}, \bibinfo {author} {\bibfnamefont {A.}~\bibnamefont {Gyenis}},
  \bibinfo {author} {\bibfnamefont {A.}~\bibnamefont {Di~Paolo}}, \bibinfo
  {author} {\bibfnamefont {A.~T.}\ \bibnamefont {Asfaw}}, \bibinfo {author}
  {\bibfnamefont {S.~A.}\ \bibnamefont {Lyon}}, \bibinfo {author}
  {\bibfnamefont {A.}~\bibnamefont {Blais}}, \ and\ \bibinfo {author}
  {\bibfnamefont {A.~A.}\ \bibnamefont {Houck}},\ }\href {\doibase
  10.1103/PhysRevLett.122.010504} {\bibfield  {journal} {\bibinfo  {journal}
  {Phys. Rev. Lett.}\ }\textbf {\bibinfo {volume} {122}},\ \bibinfo {pages}
  {010504} (\bibinfo {year} {2019})}\BibitemShut {NoStop}%
\bibitem [{\citenamefont {Gr{\"u}nhaupt}\ \emph {et~al.}(2019)\citenamefont
  {Gr{\"u}nhaupt}, \citenamefont {Spiecker}, \citenamefont {Gusenkova},
  \citenamefont {Maleeva}, \citenamefont {Skacel}, \citenamefont {Takmakov},
  \citenamefont {Valenti}, \citenamefont {Winkel}, \citenamefont {Rotzinger},
  \citenamefont {Wernsdorfer}, \citenamefont {Ustinov},\ and\ \citenamefont
  {Pop}}]{Grunhaupt2019}%
  \BibitemOpen
  \bibfield  {author} {\bibinfo {author} {\bibfnamefont {L.}~\bibnamefont
  {Gr{\"u}nhaupt}}, \bibinfo {author} {\bibfnamefont {M.}~\bibnamefont
  {Spiecker}}, \bibinfo {author} {\bibfnamefont {D.}~\bibnamefont {Gusenkova}},
  \bibinfo {author} {\bibfnamefont {N.}~\bibnamefont {Maleeva}}, \bibinfo
  {author} {\bibfnamefont {S.~T.}\ \bibnamefont {Skacel}}, \bibinfo {author}
  {\bibfnamefont {I.}~\bibnamefont {Takmakov}}, \bibinfo {author}
  {\bibfnamefont {F.}~\bibnamefont {Valenti}}, \bibinfo {author} {\bibfnamefont
  {P.}~\bibnamefont {Winkel}}, \bibinfo {author} {\bibfnamefont
  {H.}~\bibnamefont {Rotzinger}}, \bibinfo {author} {\bibfnamefont
  {W.}~\bibnamefont {Wernsdorfer}}, \bibinfo {author} {\bibfnamefont {A.~V.}\
  \bibnamefont {Ustinov}}, \ and\ \bibinfo {author} {\bibfnamefont {I.~M.}\
  \bibnamefont {Pop}},\ }\href {\doibase 10.1038/s41563-019-0350-3} {\bibfield
  {journal} {\bibinfo  {journal} {Nature Materials}\ }\textbf {\bibinfo
  {volume} {18}},\ \bibinfo {pages} {816} (\bibinfo {year} {2019})}\BibitemShut
  {NoStop}%
\bibitem [{\citenamefont {Niepce}\ \emph {et~al.}(2019)\citenamefont {Niepce},
  \citenamefont {Burnett},\ and\ \citenamefont {Bylander}}]{Niepce2019}%
  \BibitemOpen
  \bibfield  {author} {\bibinfo {author} {\bibfnamefont {D.}~\bibnamefont
  {Niepce}}, \bibinfo {author} {\bibfnamefont {J.}~\bibnamefont {Burnett}}, \
  and\ \bibinfo {author} {\bibfnamefont {J.}~\bibnamefont {Bylander}},\ }\href
  {\doibase 10.1103/PhysRevApplied.11.044014} {\bibfield  {journal} {\bibinfo
  {journal} {Phys. Rev. Applied}\ }\textbf {\bibinfo {volume} {11}},\ \bibinfo
  {pages} {044014} (\bibinfo {year} {2019})}\BibitemShut {NoStop}%
\bibitem [{\citenamefont {Matveev}\ \emph {et~al.}(2002)\citenamefont
  {Matveev}, \citenamefont {Larkin},\ and\ \citenamefont
  {Glazman}}]{matveevPRL02}%
  \BibitemOpen
  \bibfield  {author} {\bibinfo {author} {\bibfnamefont {K.~A.}\ \bibnamefont
  {Matveev}}, \bibinfo {author} {\bibfnamefont {A.~I.}\ \bibnamefont {Larkin}},
  \ and\ \bibinfo {author} {\bibfnamefont {L.~I.}\ \bibnamefont {Glazman}},\
  }\href {\doibase 10.1103/PhysRevLett.89.096802} {\bibfield  {journal}
  {\bibinfo  {journal} {Phys. Rev. Lett.}\ }\textbf {\bibinfo {volume} {89}},\
  \bibinfo {pages} {096802} (\bibinfo {year} {2002})}\BibitemShut {NoStop}%
\bibitem [{\citenamefont {Maleeva}\ \emph {et~al.}(2018)\citenamefont
  {Maleeva}, \citenamefont {Gr{\"u}nhaupt}, \citenamefont {Klein},
  \citenamefont {Levy-Bertrand}, \citenamefont {Dupre}, \citenamefont {Calvo},
  \citenamefont {Valenti}, \citenamefont {Winkel}, \citenamefont {Friedrich},
  \citenamefont {Wernsdorfer}, \citenamefont {Ustinov}, \citenamefont
  {Rotzinger}, \citenamefont {Monfardini}, \citenamefont {Fistul},\ and\
  \citenamefont {Pop}}]{Maleeva2018}%
  \BibitemOpen
  \bibfield  {author} {\bibinfo {author} {\bibfnamefont {N.}~\bibnamefont
  {Maleeva}}, \bibinfo {author} {\bibfnamefont {L.}~\bibnamefont
  {Gr{\"u}nhaupt}}, \bibinfo {author} {\bibfnamefont {T.}~\bibnamefont
  {Klein}}, \bibinfo {author} {\bibfnamefont {F.}~\bibnamefont
  {Levy-Bertrand}}, \bibinfo {author} {\bibfnamefont {O.}~\bibnamefont
  {Dupre}}, \bibinfo {author} {\bibfnamefont {M.}~\bibnamefont {Calvo}},
  \bibinfo {author} {\bibfnamefont {F.}~\bibnamefont {Valenti}}, \bibinfo
  {author} {\bibfnamefont {P.}~\bibnamefont {Winkel}}, \bibinfo {author}
  {\bibfnamefont {F.}~\bibnamefont {Friedrich}}, \bibinfo {author}
  {\bibfnamefont {W.}~\bibnamefont {Wernsdorfer}}, \bibinfo {author}
  {\bibfnamefont {A.~V.}\ \bibnamefont {Ustinov}}, \bibinfo {author}
  {\bibfnamefont {H.}~\bibnamefont {Rotzinger}}, \bibinfo {author}
  {\bibfnamefont {A.}~\bibnamefont {Monfardini}}, \bibinfo {author}
  {\bibfnamefont {M.~V.}\ \bibnamefont {Fistul}}, \ and\ \bibinfo {author}
  {\bibfnamefont {I.~M.}\ \bibnamefont {Pop}},\ }\href {\doibase
  10.1038/s41467-018-06386-9} {\bibfield  {journal} {\bibinfo  {journal}
  {Nature Communications}\ }\textbf {\bibinfo {volume} {9}},\ \bibinfo {pages}
  {3889} (\bibinfo {year} {2018})}\BibitemShut {NoStop}%
\bibitem [{\citenamefont {Manucharyan}\ \emph {et~al.}(2012)\citenamefont
  {Manucharyan}, \citenamefont {Masluk}, \citenamefont {Kamal}, \citenamefont
  {Koch}, \citenamefont {Glazman},\ and\ \citenamefont
  {Devoret}}]{ManucharyanPRB2012}%
  \BibitemOpen
  \bibfield  {author} {\bibinfo {author} {\bibfnamefont {V.~E.}\ \bibnamefont
  {Manucharyan}}, \bibinfo {author} {\bibfnamefont {N.~A.}\ \bibnamefont
  {Masluk}}, \bibinfo {author} {\bibfnamefont {A.}~\bibnamefont {Kamal}},
  \bibinfo {author} {\bibfnamefont {J.}~\bibnamefont {Koch}}, \bibinfo {author}
  {\bibfnamefont {L.~I.}\ \bibnamefont {Glazman}}, \ and\ \bibinfo {author}
  {\bibfnamefont {M.~H.}\ \bibnamefont {Devoret}},\ }\href {\doibase
  10.1103/PhysRevB.85.024521} {\bibfield  {journal} {\bibinfo  {journal} {Phys.
  Rev. B}\ }\textbf {\bibinfo {volume} {85}},\ \bibinfo {pages} {024521}
  (\bibinfo {year} {2012})}\BibitemShut {NoStop}%
\bibitem [{\citenamefont {Devoret}(1997)}]{Devoret1997}%
  \BibitemOpen
  \bibfield  {author} {\bibinfo {author} {\bibfnamefont {M.~H.}\ \bibnamefont
  {Devoret}},\ }\enquote {\bibinfo {title} {Quantum fluctuations in electrical
  circuits},}\ in\ \href@noop {} {\emph {\bibinfo {booktitle} {Quantum
  Fluctuations}}},\ \bibinfo {editor} {edited by\ \bibinfo {editor}
  {\bibfnamefont {S.}~\bibnamefont {Reynaud}}, \bibinfo {editor} {\bibfnamefont
  {E.}~\bibnamefont {Giacobino}}, \ and\ \bibinfo {editor} {\bibfnamefont
  {J.}~\bibnamefont {Zinn-Justin}}}\ (\bibinfo  {publisher} {Elsevier},\
  \bibinfo {address} {New York},\ \bibinfo {year} {1997})\ Chap.~\bibinfo
  {chapter} {10}, pp.\ \bibinfo {pages} {351--385}\BibitemShut {NoStop}%
\bibitem [{\citenamefont {Caldeira}\ and\ \citenamefont
  {Leggett}(1983)}]{Caldeira1983}%
  \BibitemOpen
  \bibfield  {author} {\bibinfo {author} {\bibfnamefont {A.~O.}\ \bibnamefont
  {Caldeira}}\ and\ \bibinfo {author} {\bibfnamefont {A.~J.}\ \bibnamefont
  {Leggett}},\ }\href@noop {} {\bibfield  {journal} {\bibinfo  {journal}
  {Annals of Physics}\ }\textbf {\bibinfo {volume} {149}},\ \bibinfo {pages}
  {374} (\bibinfo {year} {1983})}\BibitemShut {NoStop}%
\bibitem [{\citenamefont {Ferguson}\ \emph {et~al.}(2013)\citenamefont
  {Ferguson}, \citenamefont {Houck},\ and\ \citenamefont
  {Koch}}]{FergusonPRX2013}%
  \BibitemOpen
  \bibfield  {author} {\bibinfo {author} {\bibfnamefont {D.~G.}\ \bibnamefont
  {Ferguson}}, \bibinfo {author} {\bibfnamefont {A.~A.}\ \bibnamefont {Houck}},
  \ and\ \bibinfo {author} {\bibfnamefont {J.}~\bibnamefont {Koch}},\ }\href
  {\doibase 10.1103/PhysRevX.3.011003} {\bibfield  {journal} {\bibinfo
  {journal} {Phys. Rev. X}\ }\textbf {\bibinfo {volume} {3}},\ \bibinfo {pages}
  {011003} (\bibinfo {year} {2013})}\BibitemShut {NoStop}%
\bibitem [{Note1()}]{Note1}%
  \BibitemOpen
  \bibinfo {note} {To carefully verify that $\Psi _0$ satisfies the correct
  boundary conditions, regard the new variables $\theta , \theta _2, \protect
  \dots , \theta _N$ as functions of the original variables $\Theta _i$. Add $2
  \pi $ to any $\Theta _i$, and replace the index $k_i$ everywhere with
  $k^\prime _i -1= k_i$; $\Psi _0$ returns to itself.}\BibitemShut {Stop}%
\bibitem [{Note2()}]{Note2}%
  \BibitemOpen
  \bibinfo {note} {Note $q = C_d^a \DOTSB \sum@ \slimits@ _{k=1}^{N-1}
  (k/N-1/2) V_k + C_d^b(V_N-V_0)/2$. This is unchanged, as to be expected, by a
  constant shift of all $V_i$.}\BibitemShut {Stop}%
\bibitem [{\citenamefont {{Nguyen}}\ \emph {et~al.}()\citenamefont {{Nguyen}},
  \citenamefont {{Lin}}, \citenamefont {{Somoroff}}, \citenamefont {{Mencia}},
  \citenamefont {{Grabon}},\ and\ \citenamefont {{Manucharyan}}}]{Nguyen2018}%
  \BibitemOpen
  \bibfield  {author} {\bibinfo {author} {\bibfnamefont {L.~B.}\ \bibnamefont
  {{Nguyen}}}, \bibinfo {author} {\bibfnamefont {Y.-H.}\ \bibnamefont {{Lin}}},
  \bibinfo {author} {\bibfnamefont {A.}~\bibnamefont {{Somoroff}}}, \bibinfo
  {author} {\bibfnamefont {R.}~\bibnamefont {{Mencia}}}, \bibinfo {author}
  {\bibfnamefont {N.}~\bibnamefont {{Grabon}}}, \ and\ \bibinfo {author}
  {\bibfnamefont {V.~E.}\ \bibnamefont {{Manucharyan}}},\ }\href@noop {} {\
  }\Eprint {http://arxiv.org/abs/1810.11006} {arXiv:1810.11006} \BibitemShut
  {NoStop}%
\bibitem [{\citenamefont {Zorin}\ \emph {et~al.}(1996)\citenamefont {Zorin},
  \citenamefont {Ahlers}, \citenamefont {Niemeyer}, \citenamefont {Weimann},
  \citenamefont {Wolf}, \citenamefont {Krupenin},\ and\ \citenamefont
  {Lotkhov}}]{Zorin1996}%
  \BibitemOpen
  \bibfield  {author} {\bibinfo {author} {\bibfnamefont {A.~B.}\ \bibnamefont
  {Zorin}}, \bibinfo {author} {\bibfnamefont {F.-J.}\ \bibnamefont {Ahlers}},
  \bibinfo {author} {\bibfnamefont {J.}~\bibnamefont {Niemeyer}}, \bibinfo
  {author} {\bibfnamefont {T.}~\bibnamefont {Weimann}}, \bibinfo {author}
  {\bibfnamefont {H.}~\bibnamefont {Wolf}}, \bibinfo {author} {\bibfnamefont
  {V.~A.}\ \bibnamefont {Krupenin}}, \ and\ \bibinfo {author} {\bibfnamefont
  {S.~V.}\ \bibnamefont {Lotkhov}},\ }\href {\doibase
  10.1103/PhysRevB.53.13682} {\bibfield  {journal} {\bibinfo  {journal} {Phys.
  Rev. B}\ }\textbf {\bibinfo {volume} {53}},\ \bibinfo {pages} {13682}
  (\bibinfo {year} {1996})}\BibitemShut {NoStop}%
\bibitem [{\citenamefont {Krantz}\ \emph {et~al.}(2019)\citenamefont {Krantz},
  \citenamefont {Kjaergaard}, \citenamefont {Yan}, \citenamefont {Orlando},
  \citenamefont {Gustavsson},\ and\ \citenamefont {Oliver}}]{Krantz2019}%
  \BibitemOpen
  \bibfield  {author} {\bibinfo {author} {\bibfnamefont {P.}~\bibnamefont
  {Krantz}}, \bibinfo {author} {\bibfnamefont {M.}~\bibnamefont {Kjaergaard}},
  \bibinfo {author} {\bibfnamefont {F.}~\bibnamefont {Yan}}, \bibinfo {author}
  {\bibfnamefont {T.~P.}\ \bibnamefont {Orlando}}, \bibinfo {author}
  {\bibfnamefont {S.}~\bibnamefont {Gustavsson}}, \ and\ \bibinfo {author}
  {\bibfnamefont {W.~D.}\ \bibnamefont {Oliver}},\ }\href {\doibase
  10.1063/1.5089550} {\bibfield  {journal} {\bibinfo  {journal} {Applied
  Physics Reviews}\ }\textbf {\bibinfo {volume} {6}},\ \bibinfo {pages}
  {021318} (\bibinfo {year} {2019})}\BibitemShut {NoStop}%
\bibitem [{Note3()}]{Note3}%
  \BibitemOpen
  \bibinfo {note} {In each of the exponentials of \protect \cref {Eq:epsilon0},
  but not the prefactor, $E_L$ gets multiplied by $\lambda ^2$. The term
  $\protect \frac {\pi ^2}{2} \left ( 1 - \protect \frac {1}{N}\right )$ also
  gets multiplied by $\lambda ^2$.}\BibitemShut {Stop}%
\bibitem [{Note4()}]{Note4}%
  \BibitemOpen
  \bibinfo {note} {It is possible to treat $\lambda $ as a variational
  parameter in \protect \cref {Eq:Psi0} and minimize the energy of $\protect
  \mathcal {H}$ to determine $\lambda $ at each $N$. The obtained $\lambda $ is
  very close to 1 except at the smallest values of $N$.}\BibitemShut {Stop}%
\bibitem [{\citenamefont {Viola}\ and\ \citenamefont
  {Catelani}(2015)}]{Viola2015}%
  \BibitemOpen
  \bibfield  {author} {\bibinfo {author} {\bibfnamefont {G.}~\bibnamefont
  {Viola}}\ and\ \bibinfo {author} {\bibfnamefont {G.}~\bibnamefont
  {Catelani}},\ }\href {\doibase 10.1103/PhysRevB.92.224511} {\bibfield
  {journal} {\bibinfo  {journal} {Phys. Rev. B}\ }\textbf {\bibinfo {volume}
  {92}},\ \bibinfo {pages} {224511} (\bibinfo {year} {2015})}\BibitemShut
  {NoStop}%
\bibitem [{\citenamefont {Catelani}\ \emph {et~al.}(2011)\citenamefont
  {Catelani}, \citenamefont {Koch}, \citenamefont {Frunzio}, \citenamefont
  {Schoelkopf}, \citenamefont {Devoret},\ and\ \citenamefont
  {Glazman}}]{Catelani2011}%
  \BibitemOpen
  \bibfield  {author} {\bibinfo {author} {\bibfnamefont {G.}~\bibnamefont
  {Catelani}}, \bibinfo {author} {\bibfnamefont {J.}~\bibnamefont {Koch}},
  \bibinfo {author} {\bibfnamefont {L.}~\bibnamefont {Frunzio}}, \bibinfo
  {author} {\bibfnamefont {R.~J.}\ \bibnamefont {Schoelkopf}}, \bibinfo
  {author} {\bibfnamefont {M.~H.}\ \bibnamefont {Devoret}}, \ and\ \bibinfo
  {author} {\bibfnamefont {L.~I.}\ \bibnamefont {Glazman}},\ }\href {\doibase
  10.1103/PhysRevLett.106.077002} {\bibfield  {journal} {\bibinfo  {journal}
  {Phys. Rev. Lett.}\ }\textbf {\bibinfo {volume} {106}},\ \bibinfo {pages}
  {077002} (\bibinfo {year} {2011})}\BibitemShut {NoStop}%
\end{thebibliography}%

\end{document}